\documentclass[aps,prl,twocolumn,preprintnumbers]{revtex4}
\usepackage{graphicx}
\usepackage{amssymb}
\usepackage{amsmath}
\usepackage{times}
\usepackage{latexsym}
\usepackage{url}
\def\beq{\begin{equation}}
\def\eeq{\end{equation}}
\def\bey{\begin{eqnarray}}
\def\eey{\end{eqnarray}}

\def\lsim{\mathrel{\raise.3ex\hbox{$<$\kern-.75em\lower1ex\hbox{$\sim$}}}}
\def\gsim{\mathrel{\raise.3ex\hbox{$>$\kern-.75em\lower1ex\hbox{$\sim$}}}}

\newcommand{\be}{\begin{equation}}
\newcommand{\ee}{\end{equation}}

\begin{document}

\title{Light neutralinos with large scattering cross sections in the minimal supersymmetric standard model}
\author{Eric Kuflik, Aaron Pierce, and Kathryn M. Zurek}
\address{Michigan Center for Theoretical Physics, University of Michigan, Ann Arbor, MI 48109
}

\date{\today}

\begin{abstract}
Motivated by recent data from CoGeNT and the DAMA annual modulation signal, we discuss collider constraints on minimal supersymmetric standard model neutralino dark matter with mass in the 5-15 GeV range.   The lightest superpartner (LSP) would be a bino with a small Higgsino admixture.  Maximization of the dark matter-nucleon scattering cross section for such a weakly interacting massive particle requires a light Higgs boson with $\tan \beta$ enhanced couplings.    Limits on the invisible width of the $Z$ boson, combined with the rare decays  $B^{\pm} \rightarrow \tau \nu$, and the ratio $B \rightarrow D \tau \nu/B \rightarrow D \ell \nu $, constrain  cross sections to be below $\sigma_n \lesssim 5 \times 10^{-42} \mbox{ cm}^2$.  This indicates a higher local Dark Matter density than is usually assumed by a factor of  roughly six would be necessary to explain the CoGeNT excess.  This scenario also requires a light charged Higgs boson, which can give substantial contributions to rare decays such as $b \rightarrow s \gamma$  and $t \rightarrow b H^+$.  We also discuss the impact of Tevatron searches for Higgs bosons at large $\tan \beta$.
\end{abstract}
\preprint{MCTP-10-07}
\maketitle


Recently, the CoGeNT experiment has reported a signal consistent with dark matter (DM) in the mass window $7 \mbox{ GeV} \lesssim m_{DM} \lesssim 11 \mbox{ GeV}$ with a cross-section for scattering off nuclei of $3 \times 10^{-41} \mbox{ cm}^2 \lesssim \sigma \lesssim 1 \times 10^{-40} \mbox{ cm}^2$ \cite{cogent}.  While it is possible that the falling exponential observed by CoGeNT is due to a background, it is interesting that a weakly interacting massive particle (WIMP) interpretation of the low recoil energy events favors a candidate with a mass identical to that indicated by a spin-independent elastic scattering interpretation \cite{DAMA} of the annual modulation observed at the DAMA experiment  \cite{DAMAlow}.
There is tension between the DAMA/CoGeNT low mass window and the null results from XENON and the CDMS  silicon detectors at the high end of the mass window.  The tension at the lower edge of this window can be significantly reduced by an appropriate choice of the scintillation efficiency factor $L_{eff}$ \cite{Leff} and halo model, as recently discussed in \cite{CoGeNTme}.

Models which attempt to explain the closeness of the baryon and dark matter (DM) contributions to the matter density of the universe also point to a DM mass in this same range \cite{ADM}.  Models of ``WIMPless'' DM  \cite{wimpless}, singlet scalars \cite{singlet}, dark sectors connected to the visible sector by kinetic mixing \cite{kinetic}, and mirror matter \cite{mirror} also give rise to a light Weakly Interacting Massive Particle (WIMP) in a mass range consistent with the CoGeNT window.   However, before turning to such comparatively exotic scenarios, it is prudent to examine whether a more established candidate can generate such a signal.   The most studied dark matter candidate is the neutralino of the minimal supersymmetric standard model (MSSM).  Motivated by the hints from CoGeNT and DAMA, we study light MSSM DM, asking how large a cross section is achievable in these models, consistent with existing collider constraints.

The neutralino $\chi^0$ is a linear combination of bino, wino and Higgsino components, $(\tilde{B},\tilde{W},\tilde{H}_d,\tilde{H}_u)$.  In direct detection experiments, it interacts with nuclei through Higgs bosons, $Z$, and squark exchange.  In most MSSM models, the squarks tend to be heavy, limiting their effectiveness as mediators for nuclear scattering.  Their contribution is typically several orders of magnitude beneath the largest cross sections discussed here (see {\em  e.g.} \cite{CPP}).   Scattering through the $Z$ contributes spin-dependent scattering, but in light neutralino scenarios such as is relevant for the light WIMP window, the coupling to the $Z$ is limited by the invisible $Z$ width.  For large scattering cross sections in the light window, couplings to Higgs bosons dominate.

There is previous work on explaining the DAMA signal from a light MSSM LSP \cite{Bottino:2002ry} (including a discussion of the relic density), and constraining a light neutralino in the MSSM in general \cite{OldLightMSSM,Belanger:2003wb,Dreiner:2009ic}.  In this paper we revisit the light MSSM LSP in light of the recent result from CoGeNT, apply recent relevant particle physics constraints, and discuss implications for other Higgs boson mediated processes.  This region with largest scattering cross section has become constrained by Tevatron searches for MSSM Higgs bosons, particularly in the  $ \tau^+ \tau^-$ final state. The result is that a MSSM neutralino has difficulty reproducing cross sections in the CoGeNT region, but a slight overdensity of local dark matter might allow consistency.

The scattering of a WIMP with a nucleus is given by the cross section, see {\em e.g.} \cite{JKG}
\be\label{scattering-fs}
\sigma=\frac{4}{\pi}\frac{m_{\rm DM}^2m_N^2}{(m_{\rm DM}+m_N)^2}\left(Z f_p+(A-Z)f_n\right)^2
\ee
where $A$ and $Z$ are the atomic mass and atomic number of the target nuclei. The effective couplings to protons and neutrons, $f_{p,n}$, can be written in terms of the WIMP's couplings to quarks.  Since the particle which mediates the scattering is typically much heavier than the momentum transfer in the scattering, the scattering can be written in terms of an effective coupling $G_q$ :
\be \label{fpn}
f_{p,n}=\sum_{q=u,d,s} G_{q} f^{(p,n)}_{Tq}\frac{m_{p,n}}{m_q}+\frac{2}{27}f^{(p,n)}_{TG}\sum_{q=c,b,t}  G_{q}\frac{m_{p,n}}{m_q},
\ee
where $G_q=\lambda_{\rm DM}\lambda_q / M^2_M$. Here $M$ denotes the mediator, and $\lambda_{\rm DM}$, $\lambda_{f}$ denote the mediator's couplings to DM and quark.
If the mediator is a scalar Higgs boson, the $\lambda_f$ are simply the Yukawa couplings of the quarks, $y_q$, and for the $f^{p,n}_{Tq}$ we take $f_u^p=0.020,~f_d^p = 0.026,~f_s^p=0.118,~f_u^n=0.014,~f_d^n=0.036,~f_s^n=0.118$ \cite{Ellis:2000ds}.  Note the value of the strange quark content of the nucleon has a large effect on the cross section.  For example, taking the value of the strange quark content as in \cite{lattice}, as motivated by recent lattice determinations, the scattering cross sections become smaller by a factor of 2.

The neutralino masses and mixings depend on $\tan \beta=v_u/v_d$, $\mu$, and the soft gaugino masses $M_1$ and $M_2$. The scattering cross section is a function of the bino, wino and Higgsino fractions of the neutralino, decomposed as $\chi^0 = Z_B \tilde{B} + Z_W \tilde{W} + Z_d  \tilde{H}_d + Z_u \tilde{H}_u$. The masses of the lightest CP even Higgs bosons, $m_{h}$ and $m_{H}$, and the coupling of the Higgs to the quarks, as determined by $\tan \beta$ and $\alpha$, the Higgs mixing angle, are also important.  Higgsino fractions are found by diagonalizing the neutralino mass matrix. For reference, the (tree level) CP even Higgs masses are given  through the relations to the CP odd Higgs mass $m_{A}$:
\begin{eqnarray}
m_{h,H}^2 &=& \frac{1}{2}\left(m_{A}^2+m_Z^2 \right. \nonumber \\ && \left. \mp \sqrt{(m_{A}^2 - m_Z^2)^2+4 m_Z^2 m_{A}^2 \sin^2 2 \beta}\right) \nonumber \\
m_{H^\pm}^2& = &m_{A}^2+m_W^2.
\label{masses}
\end{eqnarray}

At tree level, relevant parameters for the LSP and Higgs sector phenomenology are
$\tan \beta,~~M_1,~~\mu,~~M_A,~M_2$.  Taking loop corrections into account, $~A_t$ and sfermion masses also enter.
We use Pythia 6.4 \cite{Sjostrand:2006za} to calculate spectra and branching ratios where necessary.
For large $\tan\beta$ and light Higgs region, we find the scattering cross section
\begin{eqnarray}
\sigma_n & \approx & 8.3 \times 10^{-42} \mbox{ cm}^2 \left(\frac{Z_d}{0.4} \right)^2 \left(\frac{\tan \beta}{30}\right)^2 \left(\frac{100 \mbox{ GeV}}{m_H}\right)^4 \nonumber \\
&&\times  \frac{1}{(1+\Delta m_b)^2},
\label{scatteringxsectn}
\end{eqnarray}
where we have taken the expression from \cite{Ellis:2000ds} and added important corrections from the shifts in the $b$ mass from superpartner loops, which can be ${\cal O}(1)$ at large $\tan \beta$\cite{HKR}.  These modify the Yukawa coupling as $y_{b} \rightarrow y_{b} (1+ \Delta m_{b})^{-1}$. We quantify the exact size of these corrections below.   At large $\tan \beta$,  the cross section Eq.~(\ref{scatteringxsectn}) agrees numerically with MicrOMEGAs \cite{Belanger:2008sj,Belanger:2001fz} within a few percent.  At somewhat smaller $\tan \beta$ (as will be preferred by $B$ decays, see below), this formula is good to 10\%.  We see that CoGeNT is pushing the limits of the MSSM.  To obtain a large enough scattering cross section we require a light Higgs, a substantial Higgsino fraction of the lightest neutralino, and large $\tan \beta$ to enhance the couplings of the Higgs to the nucleon.  The lighter Higgs $H$ is mostly a down type, and is nearly degenerate with the pseudoscalar Higgs $A$, as can be seen from Eq.~(\ref{masses}).  The charged Higgs also is light.   While the near exact degeneracy of the $A$ and the lighter $H$ is modified at the loop level, the correction is typically small --   in a numerical scan, covering the region  350 GeV $<M_{\tilde{f}} <$ 2 TeV, $|A|<$2 TeV, $M_3<$ 2 TeV, $|\mu|<$300 GeV,  but specializing to $20 < \tan \beta <30$, we find a maximum correction to the degeneracy  no larger than 5\%.   Similarly, the tree level relation between the pseudoscalar and charged Higgs mass is a good approximation,  with a maximum correction of 5\%.  It is often much smaller.

Since the Higgsino fraction of the neutralino should be large to maximize the cross section, constraints from the invisible $Z$ width are important. We impose the 2$\sigma$ constraint, $\Gamma(Z \rightarrow \chi^0 \chi^0) \lesssim 3$ MeV \cite{PDG}:
\be
\Gamma(Z \rightarrow \chi^0 \chi^0) = \frac{g^2}{4 \pi}  \frac{(Z_u^2 - Z_d^2)^2}{24 c^2_w} M_Z \left[ 1 -  \left(\frac{2 m_{\chi^0}}{m_Z}\right)^2\right]^{3/2}.
\ee
where $c_w$ is the cosine of the weak mixing angle. This implies a constraint, $|Z_u^2 - Z_d^2| \lesssim 0.13$.  While the scattering cross section is not directly proportional to this combination, when combined with the structure of the neutralino mass matrix, it effectively implies a limit on $Z_d^{2}$ of 0.13.  Cancellation between $Z_u$ and $Z_d$, which could allow $Z_d$ to be larger and consistent with this constraint, occurs for small $\tan \beta$.  For $M_1 \ll M_Z,M_2 $, the $Z_{d}$ bound implies $|\mu | \gsim 108 \mbox{ GeV}$.

Because the Higgs parameters are well-specified (low $m_{A^0},~m_{H^0},~ m_{H^+}$ and large $\tan \beta$), it is possible to identify several constraints.    See \cite{CarenaMenonWagner} for a recent summary of similar issues.    Both direct production of the Higgs bosons and rare decays are relevant.

First, the lightness of the charged Higgs opens the channel $t \rightarrow H^+ b$.  At tree level, and for moderate ( $\gsim 15$) $\tan \beta$, to good approximation, the width is
\be
\Gamma^{tree}(t \rightarrow b H^+) = \frac{g^2 m_t}{64 \pi M_W^2}\left(1-\frac{m_{H^{+}}^2}{m_t^2}\right)^2 m_b^2 \tan^2 \beta,
\ee
where $m_b$ should be evaluated at the top mass, $m_{b}(m_{t}) \approx 2.9$ GeV. The corrections to the $b$-quark mass, $\Delta m_{b}$, change the effective coupling of the charged Higgs (see {\em e.g.} \cite{CarenaChargedHiggs}):
\be
\Gamma^{eff}(t \rightarrow b H^+) = \frac{1}{(1+\Delta m_b)^2} \Gamma^{tree}(t \rightarrow b H^+),
\ee
We now quantify the size of the shift \cite{HKR}:
\be
\Delta m_b =( \epsilon_0 + y_t^{2} \epsilon_Y) \tan \beta,
\ee
with
\begin{eqnarray}
\epsilon_0&  = & \frac{2 \alpha_s}{3 \pi} M_3 \mu C_0(m_{\tilde{b}_1}^2,m_{\tilde{b}_2}^2,M_3^2) \\
\epsilon_Y & =& \frac{1}{16 \pi^2} A_t \mu C_0(m_{\tilde{t}_1}^2,m_{\tilde{t}_2}^2,\mu^2),
\end{eqnarray}
where
\be
C_0(x,y,z) = \frac{y \log(y/x)}{(x-y)(z-y)}+ \frac{z \log(z/x)}{(x-z)(y-z)}.
\ee

It is possible to get good estimates for the experimentally allowed ranges of $\epsilon_Y$ and $\epsilon_0$.  The limits from CDF, $BR(B_{s} \rightarrow \mu^{+} \mu^{-}) < 4.3 \times 10^{-8}$ \cite{CDFmumu}, provide an effective bound on the size of $\epsilon_{Y}$.  Following \cite{Buras:2002vd}, we have
\begin{eqnarray}
BR (B_{s} \rightarrow \mu \mu) &=& 3.5 \times 10^{-5}  \left(\frac{\tan \beta}{50}\right)^6  \left(\frac{m_t}{M_A}\right)^{4}	\nonumber \\
&\times& \frac{(16 \pi^{2}  \epsilon_Y) ^{2}}{(1 + \Delta m_{b})^2(1 + \epsilon_{0} \tan \beta)^2}.
\end{eqnarray}
This bound imposes that $\epsilon_{Y}$ make a negligible contribution to $\Delta m_{b}$ both for the charged Higgs limits above, and for additional limits below.  This bound also indicates small $|A_{t}|$, which can difficult to achieve (because of the renormalization group flow \cite{Feldman:2010ke}).  Thus, the dominant correction to the $b$ mass comes through $\epsilon_0$, the contribution from the gluino diagram.  Using sbottom masses near their Tevatron lower bounds, $m_{\tilde{b}} = 250 \mbox{ GeV}$ \cite{TeVsbottom}, $\mu = 110$ GeV, and varying the gluino mass to maximize $\epsilon_0$, we define an $\epsilon_{max} = 6 \times 10^{-3}$, which represents the largest expected value for $\epsilon_0$.  Depending on the relative sign of $\mu$ and the gluino mass, it can take either sign.

\begin{figure}
\includegraphics[width=3.0in]{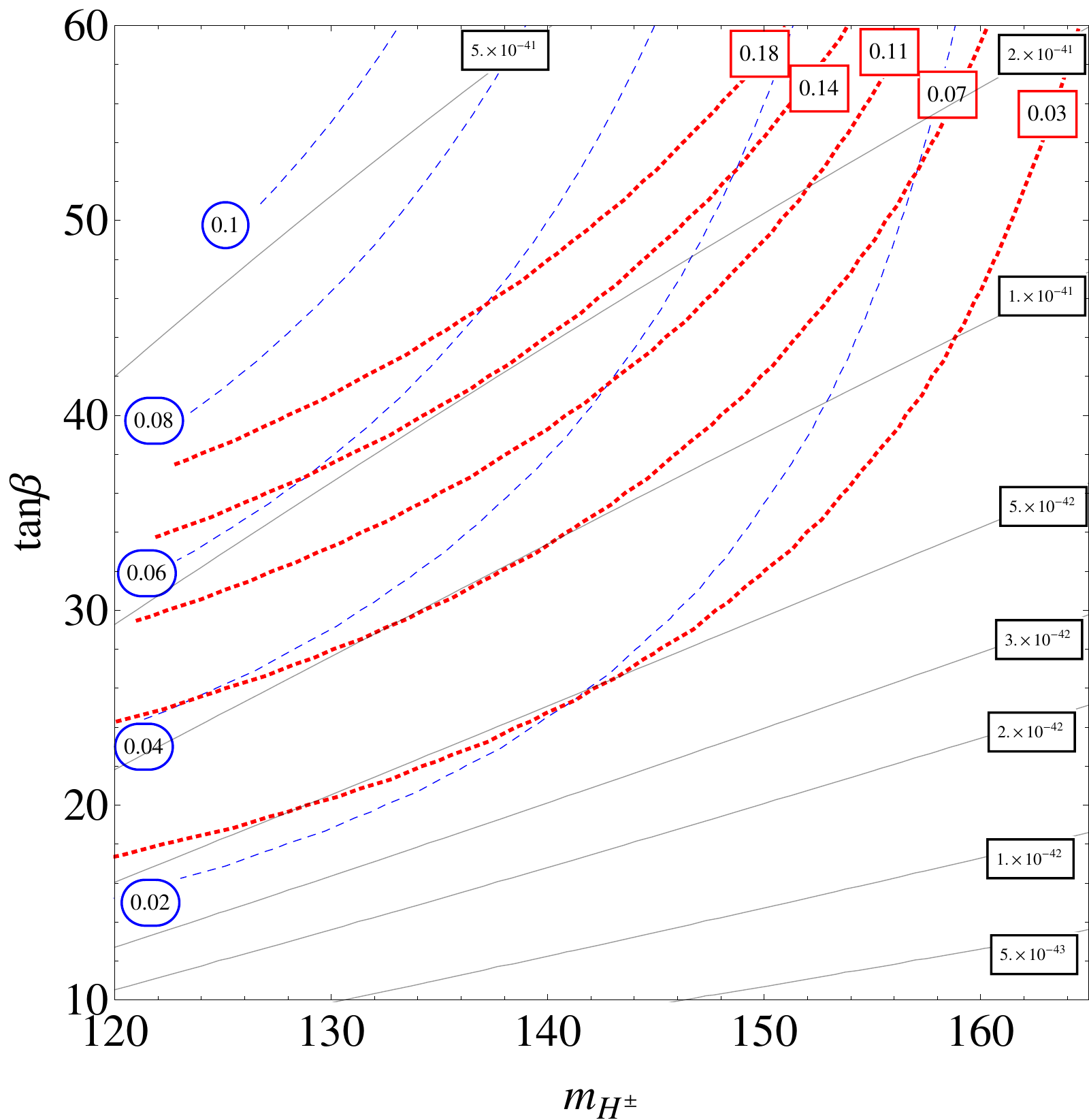}
\caption{Constraints from $t \rightarrow b H^+$ in the $m_{H^\pm}-\tan\beta$ plane.  The black solid lines indicate lines of constant scattering cross section, assuming the limit on the Higgsino fraction from the invisible $Z$ width is saturated. The dashed blue lines show the limits from $t \rightarrow b H^+$ for various branching ratios (labeled accordingly on the blue dashed lines), assuming $\epsilon_Y = 0$ and $\epsilon_0 = + 6 \times 10^{-3}$. The dotted red lines show the $\epsilon_0 = - 6 \times 10^{-3}$ limits.}
\label{fig:chargedHiggs}
\end{figure}

We show contours of the $t \rightarrow H^{+} b$ branching ratio superimposed with contours of DM-nucleon scattering cross section in Fig.~\ref{fig:chargedHiggs}.  For these cross section contours, we saturate the constraint of the invisible $Z$ width on the Higgsino fraction, and neglect the 10\% splitting between $m_{H^{\pm}}$ and $\sqrt{m_{A}^2 +m_W^2}$. For the numerical calculations of the branching ratio in the figure, we set for illustration $\epsilon_Y =0$ and $\epsilon_0 = \pm 6 \times 10^{-3}$.   The current constraints on the branching fraction from the Tevatron are vary as a function of charged Higgs mass in a range between 14\% to  25\% \cite{d0}.   If nature has chosen to live near the boundary of the currently allowed region, this decay might  be eventually observable.  It is anticipated that the LHC experiments will improve constraints on this branching ratio to $\sim 3\%$ \cite{tdr}.

At large $\tan \beta$, the charged Higgs makes a substantial contribution to the  decays $B^{\pm} \rightarrow \tau^{\pm} \nu_\tau$ and $B \rightarrow D \tau \nu$.
Taken together, measurements of these branching ratios place a strong constraint on two Higgs doublet models such as the MSSM.  First, we consider  $B^{\pm} \rightarrow \tau^{\pm} \nu_\tau$.  The ratio of the MSSM to SM expectation is ({\em e.g.} \cite{CarenaMenonWagner})
\begin{eqnarray}
R_{B \tau \nu} &= & \frac{BR(B \rightarrow \tau \nu)_{MSSM}}{BR(B \rightarrow \tau \nu)_{SM}} \nonumber \\ &=& \left[ 1 - \left( \frac{m_B^2}{m_{H^\pm}^2}\right) \frac{\tan^2\beta}{1+\epsilon_0 \tan \beta}\right]^2,
\end{eqnarray}
where $\epsilon_0$ is defined as above.  Note the charged Higgs interferes destructively with the SM contribution.  Thus, a contribution from the charged Higgs can make the branching ratio too small.  Alternately, if the charged Higgs overwhelms the SM contribution, it can give too large a rate.
There are tentative observations from BaBar and Belle, with combined significance of greater than $4\sigma$ deviation from zero, with a central value within approximately two standard deviations of the Standard Model expectation \cite{RosnerStone}. Using the SM predicted value $BR(B \rightarrow \tau \nu)_{SM} = (0.98 \pm 0.24) \times 10^{-4}$ \cite{UTFIT}, and the combined experimental observation, $BR(B \rightarrow \tau \nu)_{SM} = (1.73 \pm 0.34) \times 10^{-4}$, we find $R_{B \tau \nu} = 1.77 \pm 0.55$.  The corrections due to nonzero $\epsilon_0$ are significant at large $\tan \beta$.
Next, we turn to the process $B \rightarrow D \tau \nu$, where again, the charge Higgs boson can make a substantial contribution. Following \cite{Talk}, we combine measurements from BaBar and Belle to find  $R_{B D \tau \nu} \equiv BR(B \rightarrow D \tau \nu ) / BR (B \rightarrow D \ell \nu) = 0.49 \pm 0.1$. Using the theory formula from \cite{BDtaunu}, we extract constraints on the charged Higgs contribution to the process.

\begin{figure}
\includegraphics[width=3.0in]{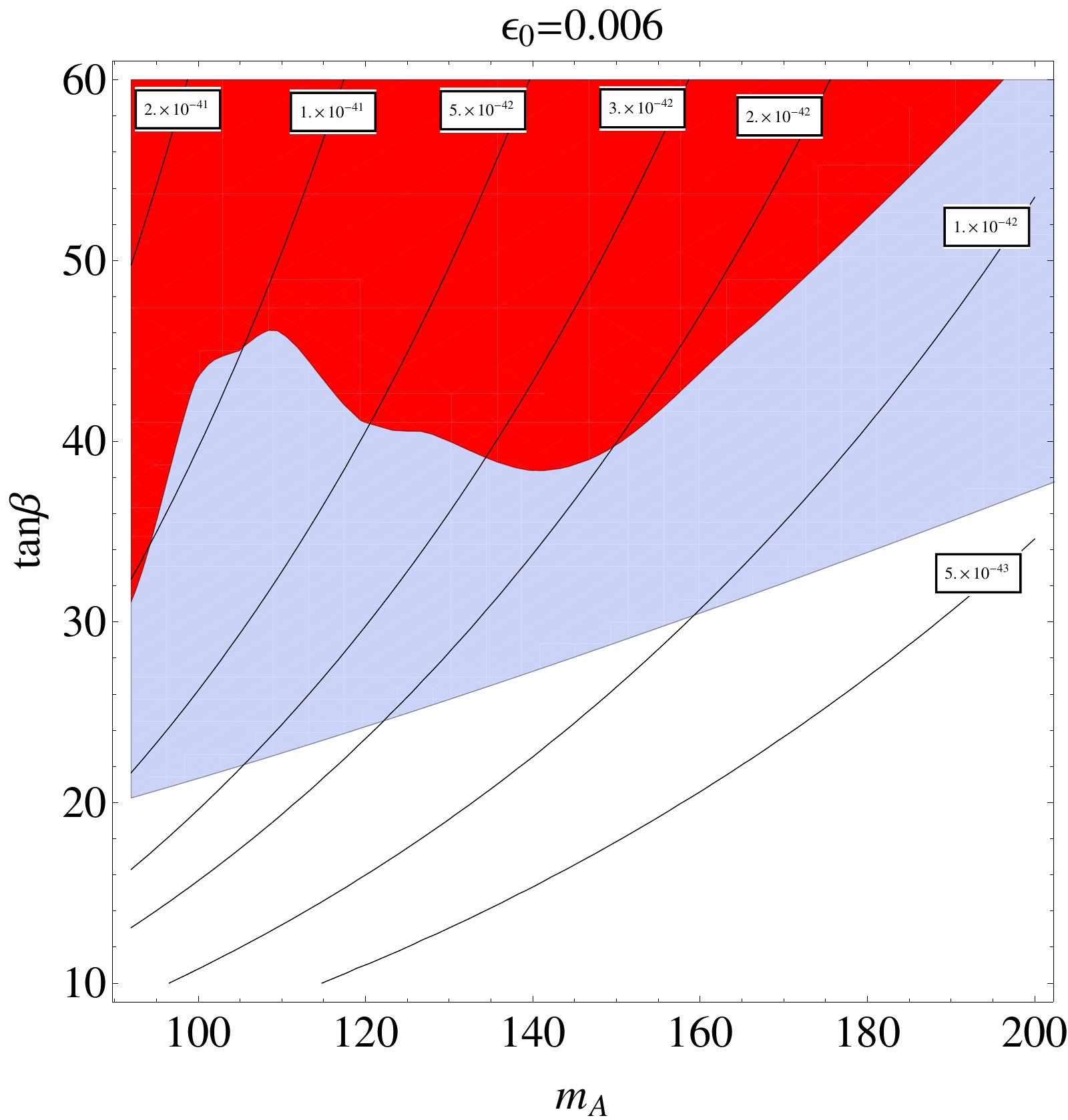}
\caption{Constraints on the $m_A-\tan\beta$ plane from $B \rightarrow \tau \nu$,  $B \rightarrow D \tau \nu$ and $\phi \rightarrow \tau^+ \tau^-$.   In the case of the $B$ decays, we show a conservative bound (grey shaded region): the intersection of the 3 sigma allowed regions for both $B$ processes. For $\phi \rightarrow \tau^+ \tau^-$  (the irregular red shaded region), the region below the curve is allowed at 2 $\sigma$ by the Tevatron.  The B-decay region depends on the squark and gluino masses due to loop corrections to the $b$ mass,  so we show  the region corresponding to $\epsilon_0=+\epsilon_{max}$.  The region for  $\epsilon_0=-\epsilon_{max}$ is shown in Fig.~\ref{fig:summary2}.  The  $\phi \rightarrow \tau^+ \tau^-$ is relatively insensitive to these corrections. We also show in this plane contours of constant scattering cross section, assuming the bound on the invisible $Z$ width (3.0 MeV) is saturated and $\epsilon_0 = + \epsilon_{max}$.}
\label{fig:summary1}
\end{figure}

\begin{figure}
\includegraphics[width=3.0in]{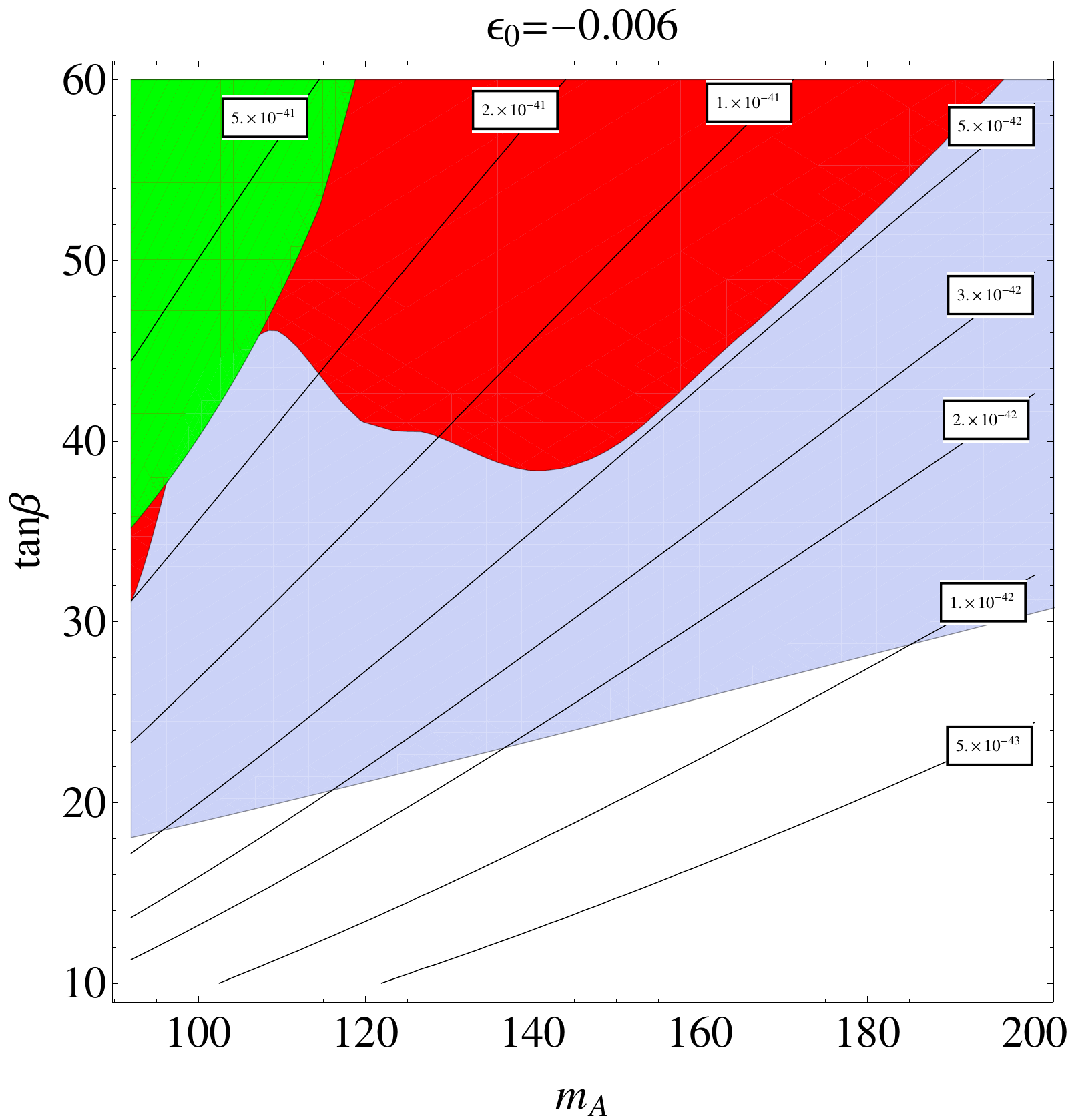}
\caption{Constraints on the $m_A-\tan\beta$ plane from $B \rightarrow \tau \nu$,  $B \rightarrow D \tau \nu$ and $\phi \rightarrow \tau^+ \tau^-$, and $t \rightarrow b H^+$.   In the case of the B decays, we show a conservative bound (grey shaded region):  the intersection of the 3 sigma allowed regions for both $B$ processes.  For $\phi \rightarrow \tau^+ \tau^-$  (the irregular red shaded region), the region below the curve is allowed at 2 $\sigma$ by the Tevatron.  Since the B-decay region depends on the squark and gluino masses due to loop corrections to the $b$ mass,  we show  lines corresponding to $\epsilon_0=-\epsilon_{max}$.   The region for  $\epsilon_0=+\epsilon_{max}$ is shown in Fig.~\ref{fig:summary1}. The  $\phi \rightarrow \tau^+ \tau^-$ constraint is relatively insensitive to these corrections.  The green shaded region indicates the constraint from $t \rightarrow b H^+$.  We also show in this plane contours of constant scattering cross section, assuming the bound on the invisible $Z$ width (3.0 MeV) is saturated and $\epsilon_0 = - \epsilon_{max}$.}
\label{fig:summary2}
\end{figure}

In Figs.~\ref{fig:summary1} and~\ref{fig:summary2}, we show the intersection of parameter space where the constraints from both the $B \rightarrow \tau \nu$ and $B \rightarrow D \tau \nu$ are both within their $3\sigma$ allowed region.  We view this as a conservative prescription.   In making these figures, we have neglected the radiative corrections to $H^\pm$ which depend on the superpartner spectrum.  However, recall these corrections are  $\lesssim 10\%$, so the effects on the direct detection cross section are 40\% at most. Figure ~\ref{fig:summary1} shows the result when $\epsilon_0 = +\epsilon_{max}$, and Fig.~\ref{fig:summary2} shows the result when $\epsilon_0 = -\epsilon_{max}$.  We have also included on these plots curves of constant scattering cross section.

In addition, tight constraints are also derived from Tevatron exclusion curves on direct scalar production at large $\tan \beta$.  Since the lightest CP even and odd Higgses are nearly degenerate in the region of parameter space relevant for light WIMPs with large scattering cross sections, we consider Tevatron constraints from both $A \rightarrow \tau^+ \tau^-$ and $H \rightarrow \tau^+ \tau^-$ \cite{Abulencia:2005kq,Owen:2007ti}.
We show the exclusions derived from these analyses in Figs.~\ref{fig:summary1},~\ref{fig:summary2} \cite{PDG} .  Because corrections from $\epsilon_0$ change the branching fraction and production cross section in opposite directions, even extreme values of $|\epsilon_0| = \epsilon_{max}$ give rise to small modifications, $\sim 5\%$, to these curves.  Examining these plots, we can pick out the largest allowed scattering cross section,  $\sigma_n \lsim 5 \times 10^{-42}$ cm$^2$, below the CoGeNT allowed region.  If the errors are both $B$ experiments are inflated even further (both experiments taken at 3.1 sigma), a fine-tuned region at larger $\tan \beta$ opens.  There the charged Higgs contribution is exactly the right size to (over)cancel the standard model contribution, such that the resulting sum is again the same size as the standard model one.  If this strip were to open, the cross allowed cross section is approximately a factor of 2 higher, $\sigma_n \lsim 1 \times 10^{-41}$ cm$^2$, and the Tevatron constraints on Higgs production would start to be relevant.

Finally, we comment on the more model-dependent flavor physics implications.  For $b \rightarrow s \gamma$,  without cancellation, such large values of $\tan \beta$ would require charged Higgs masses closer to 300 GeV \cite{GiudiceGambino}.
 In principle, there is the possibility of large canceling contributions.  However, this requires a large contribution from squark/gaugino diagrams ({\em e.g.} with light stops and charginos).    Such a delicate cancelation would be surprising, and might well show up elsewhere depending on how it were implemented ({\em e.g.}, non-minimal flavor  violation).

To conclude, acquiring a large scattering cross section in the MSSM for light WIMPs requires a very particular Higgs boson spectrum.  To achieve the largest possible cross section consistent with constraints, we require $\mu$ very near its bound at 108 GeV, sbottoms and gluino relatively light (around 350 GeV), a heavy right-handed stop around $\gsim 1.5$ TeV, and small A-terms.  To maximize scattering, the CP even Higgs boson with $\tan \beta$--enhanced couplings should be as light as possible.  At present, bounds from $B$ decays are most constraining.   Depending on the details of the SUSY spectrum, constraints from the rare decay $t \rightarrow b H^{+}$ could eventually become competitive.  We find that for WIMPs in the 5-15 GeV range, the scattering cross section must be smaller than $5 \times 10^{-42} \mbox{ cm}^2$.

Thus it appears a MSSM neutralino is in tension with the data from CoGeNT.   To explain the observed rates in these detectors would require local overdensity in the DM of a factor  of 6 to hit the edge of the window.  We leave for future work a discussion of the effect of a thermal relic history on the allowed parameter space of the low mass MSSM window, but it is interesting to note that that region near the CoGeNT window gives rise to approximately the correct relic density.\\

We thank Tim Cohen and Dan Phalen for discussions.  A.P. was supported in part by NSF Career Grant NSF-PHY-0743315 and in part by DOE Grant DE-FG02-95ER40899.  E.K. acknowledges support from NSF grant PHY/0917807 and from DOE Grant DE-FG02-95ER40899.

Note added: While this work was in preparation \cite{Feldman:2010ke} appeared which explores similar issues.

\end{document}